\documentclass[a4paper,fleqn,usenatbib]{mnras}
\usepackage{mathptmx}
\usepackage[T1]{fontenc}
\usepackage[english]{babel}
\usepackage{ae,aecompl}
\usepackage{float}% If comment this, figure moves to Page 2
\usepackage{graphicx}   % Including figure files
\usepackage{amsmath}	% Advanced maths commands
\usepackage{amssymb}	% Extra maths symbols
\usepackage{latexsym}
\usepackage{times}
\usepackage{subfig}
\usepackage{multirow}
\usepackage{array}
\usepackage{fancyhdr}
\newcommand{\be}{\begin{equation}}
\newcommand{\ee}{\end{equation}}
\title{Discovery of a shock front in the merging cluster of galaxies A2163}
\author[N. Mhlahlo et al.]{
N. Mhlahlo,$^{1}$\thanks{E-mail: nmgaxamba@gmail.com}
L. Guennou,$^{2}$ 
L. Feretti,$^{3}$ 
\\
$^{1}$School of Physics, University of the Witwatersrand, Private Bag 3, 2050-Johannesburg, South Africa  \\ 
$^{2}$University of KwaZulu-Natal, King George V Ave, Durban, 4041, South Africa \\
$^{3}$Istituto di Radioastronomia INAF, Via P. Gobetti 101, 40129 Bologna, Italy.
}
\date{Accepted XXX. Received YYY; in original form ZZZ}
\pubyear{2016}
\begin{document}
\label{firstpage}
\pagerange{\pageref{firstpage}--\pageref{lastpage}}
\maketitle

\begin{abstract}

ACO2163 is one of the hottest (mean $kT=12-15.5$ keV) and extremely X-ray overluminous merging galaxy clusters which is located at $z=0.203$. The cluster hosts one of the largest giant radio halos which are observed in most of the merging clusters, and a candidate radio relic. Recently, three merger shock fronts were detected in this cluster, explaining its extreme temperature and complex structure. Furthermore, previous {\it XMM-Newton} and {\it Chandra} observations hinted at the presence of a shock front that is associated with the gas `bullet' crossing the main cluster in the west-ward direction, and which heated the intra-cluster medium, leading to adiabatic compression of the gas behind the 'bullet'. The goal of this paper is to report on the detection of this shock front as revealed by the temperature discontinuity in the X-ray XMM-Newton image, and the edge in the Very Large Array (VLA) radio image. We also report on the detection of a relic source in the north-eastern region of the radio halo in the KAT-7 data, confirming the presence of an extended relic in this cluster.  \\
  The brightness edge in the X-rays corresponds to a shock front with a Mach number $M= 2.2\pm0.3$, at a distance of 0.2 Mpc from the cluster centre. An estimate from the luminosity jump gives $M=1.9\pm0.4$.    
We consider a simple explanation for the electrons at the shock front, and for the observed discrepancy between the average spectral index of the radio halo emission and that predicted by the $M=2.2$ shock which precedes the 'bullet'.
\end{abstract}
%----------------------------------------------------
%%%%

\section{Introduction}

Radio studies of supernovae remnants have indicated that in these sources, a fraction of the shock energy can be converted into the acceleration of the electrons to relativistic speeds \citep[e.g.][]{Blan87}. It is generally accepted that this process also takes place during cluster mergers, and is responsible for the production of the large-scale diffuse synchrotron radio emission observed in galaxy clusters' peripheries in the form of relic sources \citep[see e.g.][]{Gia08, Fin10, Mac11, Aka13, Bou13, Ogr13a}. 
The other large-scale diffuse sources that are observed at cluster centres in approximately 60 clusters so far, known as radio halos, are thought to result from the acceleration of the electrons by magnetohydrodynamic turbulence in the presence of magnetic fields, which also occurs during cluster mergers \citep[e.g.][and references therein]{Buo01, Fer02, Fer05, Fer12}.  \\
However, the spatial coincidence of shocks with the edges of radio halos, which has been observed in a few clusters \citep[e.g.][]{Mar05}, has raised questions about the role that shock acceleration processes can play in the formation of radio halos (see also the review by \citealt{Mar10}).  
Out of all the clusters where detections of merger shock fronts have been reported, roughly seven have their radio halo edges coincident with the shock location, suggesting that shocks could be responsible for (re)acceleration of relativistic electrons at the halo edge locations in the presence of magnetic fields, leading to the production of at least some of the observed radio emission at those sites.  \\
A number of models have been used to explain the electrons at the shock front. Direct acceleration of relativistic electrons by the shock \citep[e.g.][]{Blan87} has been cited as the main mechanism responsible for the emission at cluster outskirts, and at least for some of the halo emission, in a few clusters \citep{Mar05, Mac11}. Compression by the shock of fossil electrons in the intra-cluster medium (ICM) \citep[e.g.][]{Ens98, Ens01, Mar05, Mac11} has been considered to be another alternative. The compression of pre-existing electrons at the shock front is expected to significantly increase their synchrotron emission at the observing frequency and to produce radio emission in front of the bow shock \citep{Ens01}. Another process for radio emission production, which is expected to be more efficient than the alternatives, is shock re-acceleration of fossil (pre-existing) relativistic plasma which could be the remnant of a radio galaxy, by Diffusive Shock Acceleration (DSA) \citep{Blan87, Shi15, Kan16}. This model has been used to explain the observed discrepancy between the X-ray Mach number and the DSA predictions for some of the clusters \citep[e.g.][]{Kan16, Ogr13}. \\
There is no strong agreement about which mechanism best describes the electrons at the shock front. However, it is generally agreed that the sharp radio edges which are observed at the borders of some of the giant radio halos suggest a possible connection between merger shocks and the generation of turbulence in the ICM, and that acceleration by turbulence is responsible for the cluster-scale radio emission \citep[e.g.][]{Mar10, Mac11}.  \\
The evolution of cluster-scale radio emission in clusters has been investigated by \cite{Don13} who combined an idealized model for cluster mergers, with a numerical model for the injection, cooling and re-acceleration of cosmic-ray electrons. The simulations have shown that re-acceleration of cosmic-ray electrons can potentially reproduce key observables of radio halos such as the transient nature of radio halos and their connection to mergers and merger-driven turbulence.
\begin{figure*}
%\centering
  \caption{Left: Background-subtracted XMM-Newton X-ray image of the cluster ACO2163 in the 0.5 - 8.0 keV energy band, representing $\sim$9 ks of data. The image has been smoothed with a Gaussian kernel with $FWHM = 10$ arcsec. To the south-west is the brightness edge that we argue is a shock, and the two lines near the bright emission region show the approximate extent of the edge. The linear scale bar has units counts s$^{-1}$. The radio contours of the radio halo in VLA at 1.4 GHz with a resolution of 15$^{\prime\prime}$ are overlayed on the X-ray image. The $\sigma$ noise level in the image plane is 0.03 mJy/beam. Contours start at 0.09 mJy/beam and then scale by a factor of $\sqrt2$. The radio image is from \citep{Fer01}. Right: Sectors showing cluster regions where spectra were extracted both toward the south-western direction and around the edge (top) and in the perpendicular direction (bottom).}
\vspace{0.2cm}
\begin{tabular}{cc}
\centering
\multirow{2}{*}{\subfloat{\includegraphics[scale=0.7]{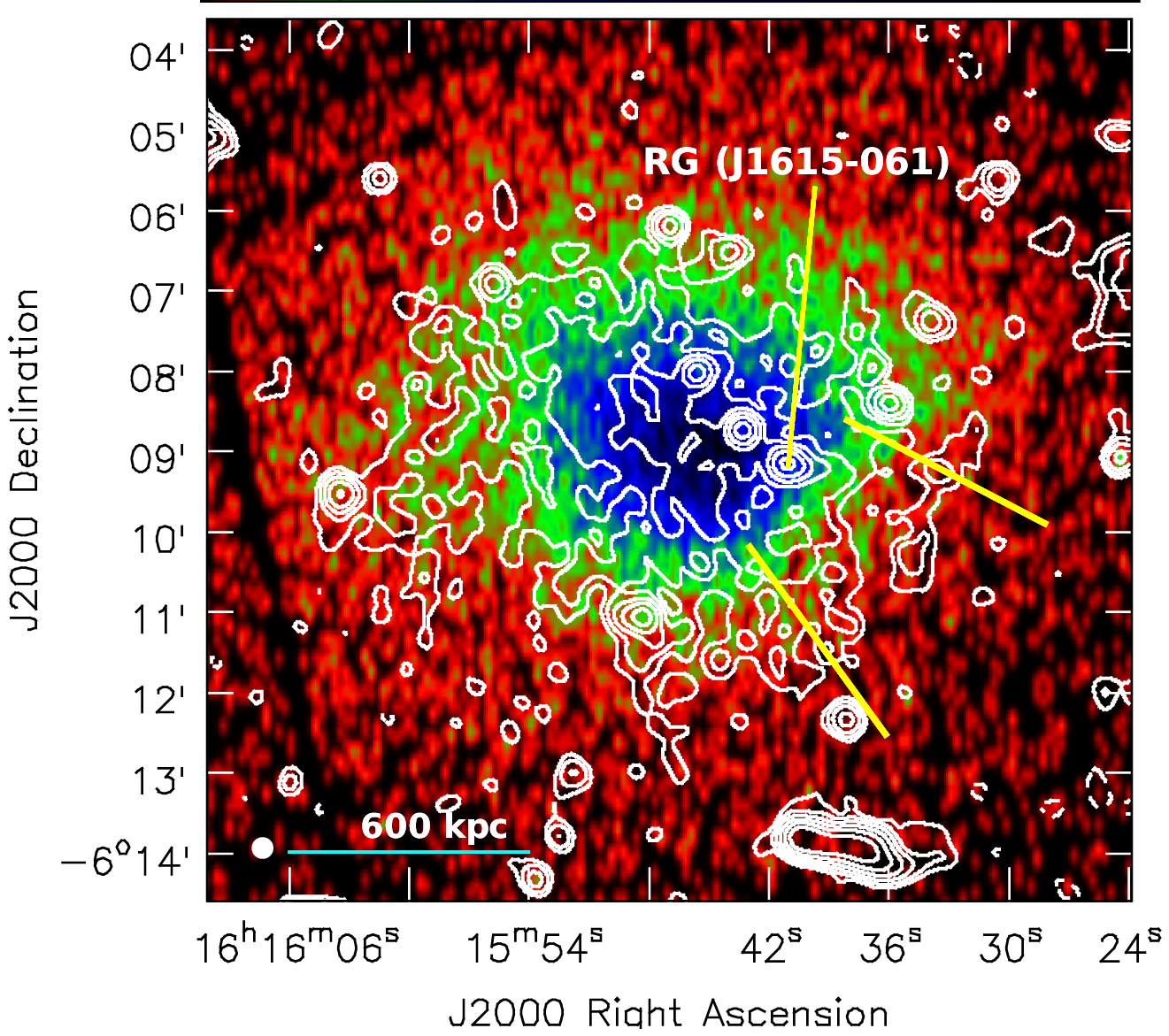}}} &
\subfloat{\includegraphics[scale= 0.3,bb=290 150 650 560, height=45mm, clip=true]{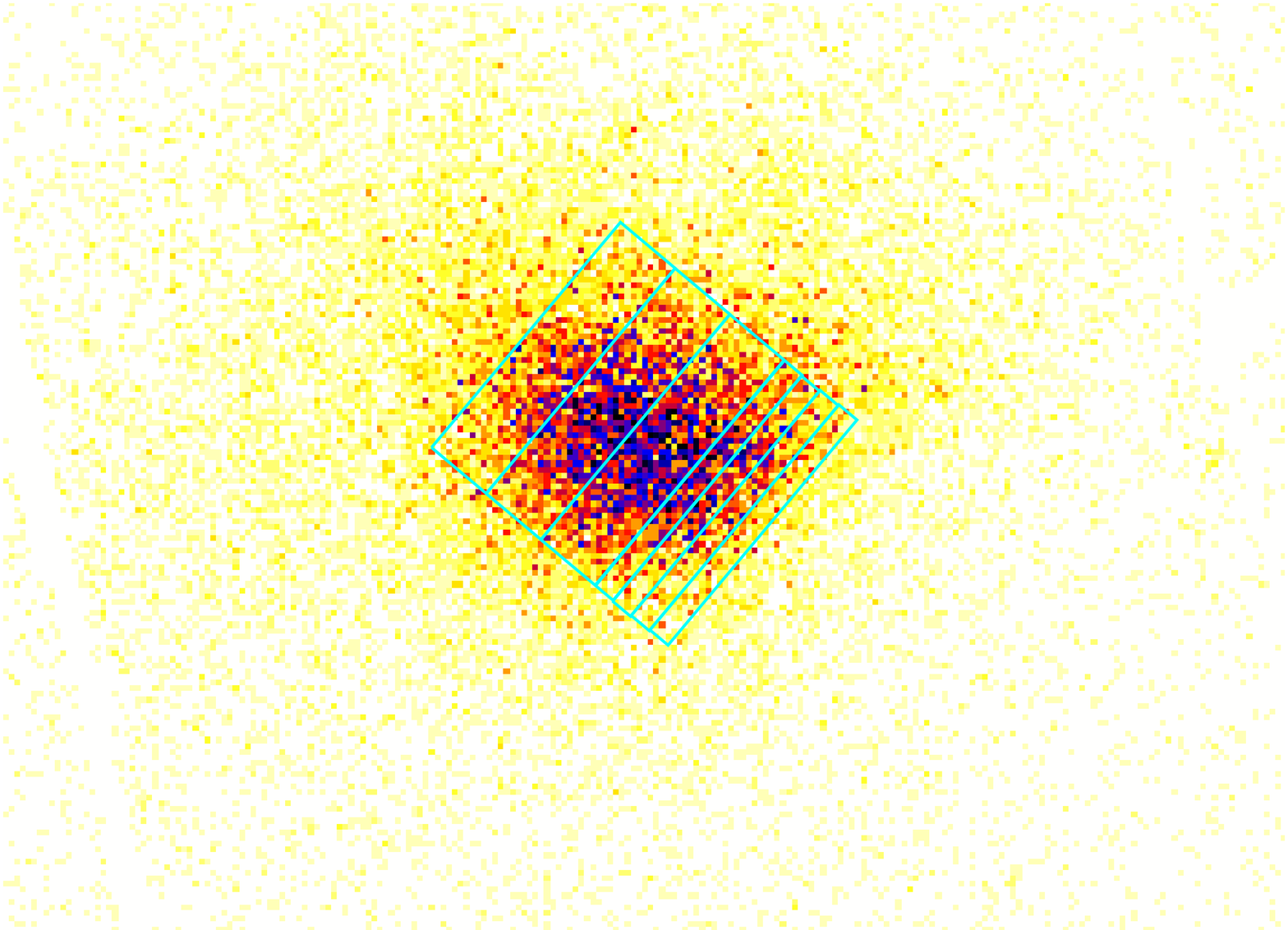}} \\ & \subfloat{\includegraphics[scale=0.5]{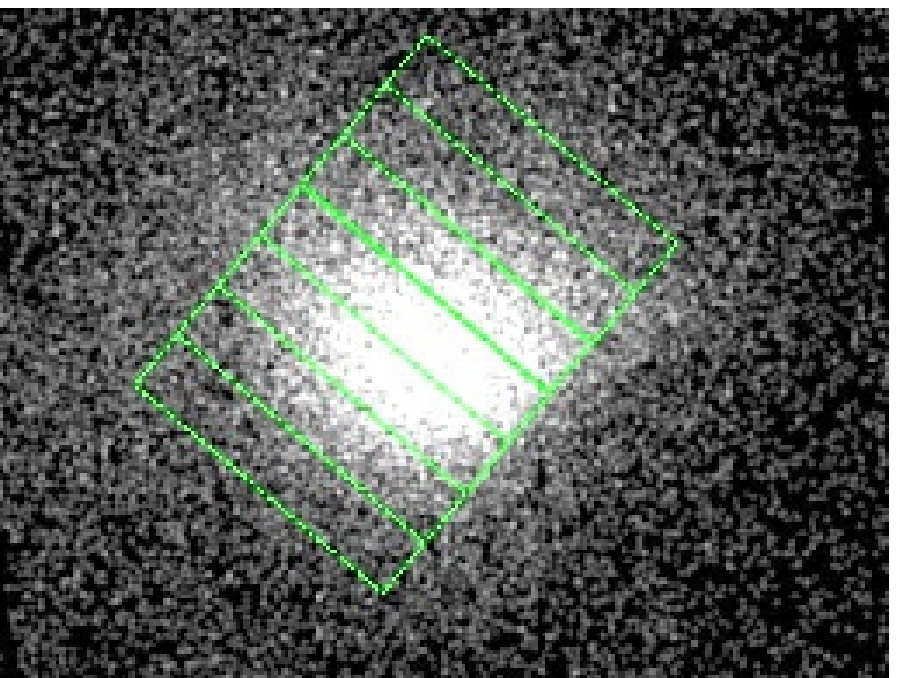}} \
\end{tabular}
\label{figa2163_xray}
\vspace{1cm}
\end{figure*}
Furthermore, non-thermal effects connected to cosmological shock waves and AGN feedback have also been studied with cosmological simulations. \cite{Vaz13} modelled the injection and evolution of cosmic rays, as well as their effects on the thermal plasma, and further investigated the injection of turbulent motions into the ICM from both the accretion of matter and AGN feedback in order to constrain the energetics and mechanisms of feedback models in clusters. Also, the simulations of \cite{Vaz11} have shown that merger shock heating is the leading source of entropy production in clusters and is responsible for generating most of the entropy of the large-scale structures in the Universe.  \\
Though there is a relatively large number of merging clusters (>60), shock fronts are rare (at a fraction of $\sim$20\%) and a discovery of just one is not only exciting but important for aiding our understanding of the physical processes occuring in the ICM. Shocks offer an opportunity to study the acceleration processes in the ICM in the presence of magnetic fields. They also enable us to investigate their connection with mergers and with the formation of the large-scale diffuse radio sources that are observed in merging clusters.   \\
In this work we report on the detection of a new shock front in the merging cluster of galaxies ACO2163. We present the analysis of X-ray XMM-Newton observations, and study the connection between the shock and the diffuse cluster radio emission, through the VLA and KAT-7 observations at 1.4 GHz and 1.83 GHz, respectively. \\
We start by discussing the cluster ACO2163 in Section~\ref{sec:a2163}. We then present the results from the X-ray observations and analysis in Section~\ref{sec:xobs}, which are followed by those at radio wavelengths in Sections~\ref{sec:robs} and \ref{sec:radio}. The discussion and conclusions are presented in Section~\ref{sec:disconc} and~\ref{sec:disconc-1}, respectively. \\
Throughout the paper a flat, vacuum-dominated Universe with $\Omega_m = 0.32$ and $\Omega_\Lambda = 0.68$ and $ H_0=67.3$ km s$^{-1}$Mpc$^{-1}$ is assumed. 
\section{The Merging Cluster ACO2163}
\label{sec:a2163}
ACO2163 is a moderately distant (z=0.203; \citealt{Str99}), rich cluster, and is one of the hottest (mean kT=12-15.5 keV, \citep{Elb95, Mar96, Mau08}) and extremely X-ray overluminous compared to its mass ($L_X$[2-10 keV]$=6\times10^{45}$erg s$^{-1}$, \citep{Arn92}). This cluster has been a subject of extensive studies at multiple wavelengths \citep[see][and references therein]{Elb95, Mar96, Squ97, Gov04, Fer01, Nor09, Bou11}. Evidence for a recent merger that involves two or more components has been revealed by X-ray morphological studies of \cite{Elb95} based on ROSAT data, the spectroscopic analysis of \cite{Mar96} based on ASCA data, and that of \cite{Bou11} based on XMM-Newton and Chandra data. In Figure~\ref{figa2163_xray} we show a smoothed X-ray image of the cluster, and we observe hints indicating the presence of a major merger, with the main axis of the cluster along the north-east, south-west direction.   \\
Dynamical studies based on joint weak gravitational lensing and X-ray observations \citep{Squ97} have shown that ACO2163 is in a disturbed dynamical state, showing irregular distribution of mass and galaxies. This has been supported by weak-lensing studies of the dark matter distribution in ACO2163 which have suggested that a multiple merger is taking place in this cluster \citep{Sou12}. Further evidence for a merger has come from X-ray temperature maps which have shown strong temperature variations and complex X-ray thermal structure in their central regions \citep{Gov04, Bou11}. Temperature maps have also revealed a cold front in the south-west direction from the cluster centre associated with $\sim$8 keV gas and surrounded by hotter $\sim$11 keV gas \citep[e.g.][]{Owe09}.  \\
The merger scenario in ACO2163 is also supported by the presence of a radio halo in this cluster. The detection of a radio halo was first reported by \cite{Her94}. \cite{Fer01} re-observed the cluster at 1.4 GHz with the VLA, and the detection of one of the largest radio halos with a total extent of $\sim$11.5$^{\prime}$ was reported \citep[see also][for low frequency observations]{Fer04}. Diffuse emission at the N-E side of the halo \citep{Fer04} was also detected and this feature was identified with a relic source.  \\
Recent studies on ACO2163 done by \cite{Bou11}, which revealed striking similarities between ACO2163 and the `Bullet' cluster, have suggested the presence of a shock front in this cluster. Evidence for a gas 'bullet' or cool-core separated from its galaxies and crossing the ACO2163 atmosphere along the east-west direction was presented. From evidence of pressure excess, particularly in the innermost regions of the main cluster ACO2163-A, \cite{Bou11} infered adiabatic compression of the gas in the ICM behind the westward moving gas `bullet' as a result of presumed shock heating. 
%
%
%\vspace{10pt}\vspace{-20pt}{p{0.6\textwidth} p{0.6\textwidth}}
%\begin{figure} 
%<<label=xcube, fig=TRUE, echo=FALSE, pdf=FALSE, eps=TRUE, png=TRUE, include=FALSE>>=
This has suggested that the main-cluster in ACO2163 has accreted a subcluster along the east-west direction at a supersonic velocity. Though the merger event might have shocked the main-cluster atmosphere, however, there was no evidence for a shock front preceding this `bullet', contrary to what is observed in the `Bullet' cluster.  \\
More recently, \cite{Thol18} detected three shock fronts in the spectral analysis of data from Suzaku XIS observations of A2163, one in the NE direction and two in the SW direction. Their inner shock in the SW direction is located at 3.3' from the cluster centre, which is further out compared to the location of the shock reported here, which is at $\sim 1'$ from the centre.
\section{X-ray Observations, Analysis and Results}
\label{sec:xobs}
The XMM-Newton observations of ACO2163 were done in August 2000. We retrieved the data of this cluster from the XMM-Newton Science Archive (obsID 0112230601, PI M. Turner, public data). The raw time of observation was 16.3 ks. The data were obtained with the THIN1 filter with the EPN and EMOS 1 and 2 cameras and then analysed using the SAS (Science Analysis System developed by the XMM-Newton team) tool from the Heasarc package to do the main part of the reduction. After removing the flares, we obtained a reduced image of $\sim$9 ks for our study. We also obtained a blank field, representing the modelisation of the background to check the possible existence of hard/soft excess that could have altered the results obtained from our data, and no excess was observed. \\
The X-ray image in Figure~\ref{figa2163_xray}, overlaid with contours of the high resolution radio image, shows a brightness edge to the south-west of the cluster with a clear cut that is observed in this direction. We have focused on this area, performing several checks on the existence of a shock. 
\begin{figure*}
\centering
\includegraphics[scale= 0.3]{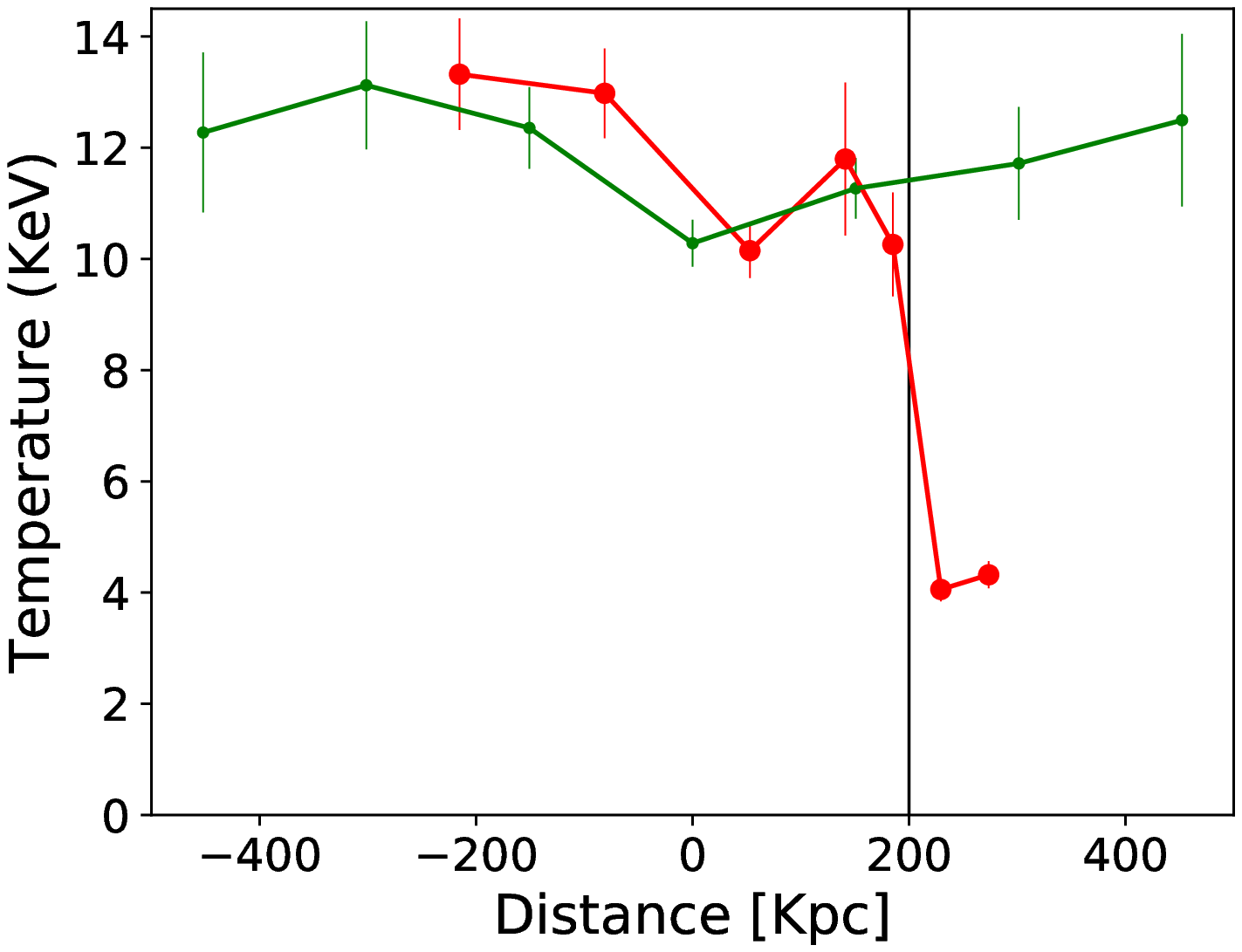}
\includegraphics[scale= 0.3]{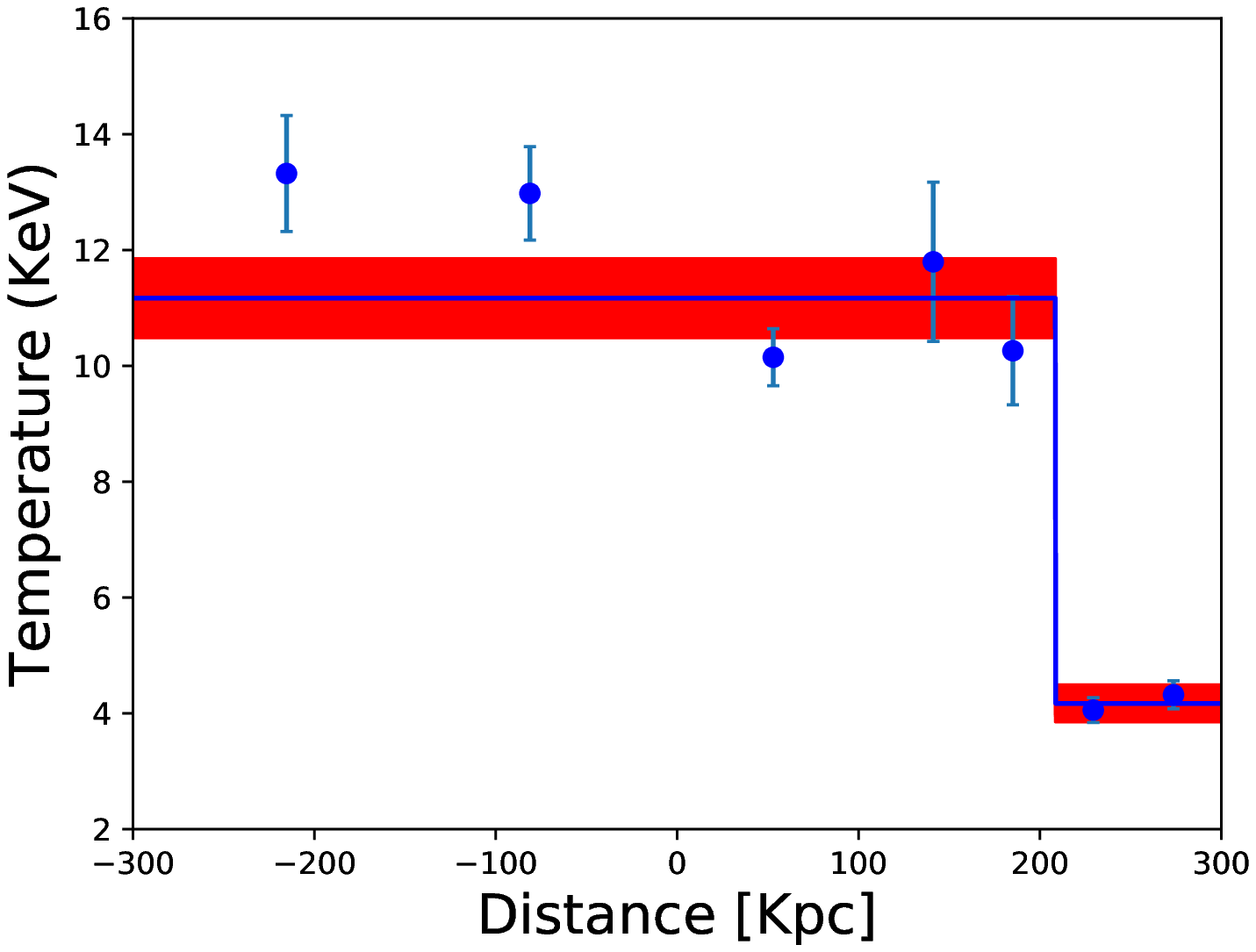}
\includegraphics[scale= 0.3]{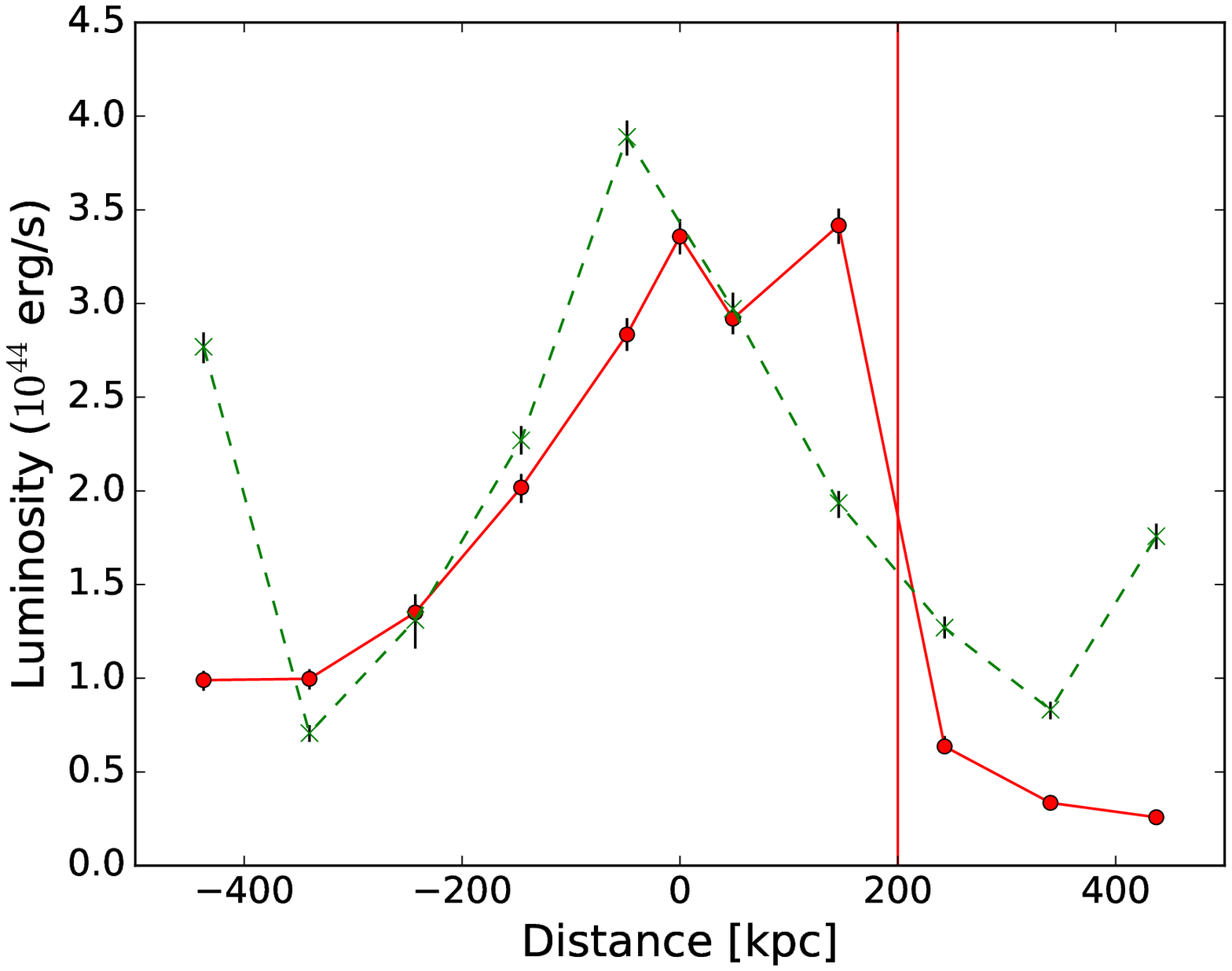}
\caption{{Left panel: Temperature profiles along the main axis (red line) and in the perpendicular direction (green line). The vertical line represents the location where the shock is expected. The null distance represents the center of the cluster. The temperature obtained in the inner regions of the cluster is similar to the ones found in the literature \citep[see e.g.][]{Mar01, Bou11}}. Middle panel: Temperature profile around the region of the shock. The shaded areas show 68\% confidence intervals. The solid line is the best-fit model (see text). Right panel: Plots of the luminosity obtained along the main axis (red dots) and in the perpendicular sector (green dots). The location of the luminosity jump is shown by the vertical line.}
\label{temp_prof}
\end{figure*}
We looked for the variations in temperature in the cluster using the combination of {\it phabs} and {\it mekal} models.  The {\it mekal} model is an emission spectrum from hot diffuse gas based on the model calculations of \citet{Mew85, Mew86} with Fe L calculations by \citet{Lie95}. The model includes line emissions from several elements, with the plasma temperature (in keV), H density (cm$^{-3}$) and Metal abundance as free parameters, to fit the different spectra taken from MOS1, MOS2 and PN, extracted in the 0.5-8.0 kev band range from several rectangular bins in the same sector across the cluster and the brightness edge.  \\
To this end, we used the sectors shown in Figure~\ref{figa2163_xray} - right panels - to extract spectra and obtain the temperature. We also extracted a background which we then subtracted from our spectra and took care of masking possible point sources in the background and cluster that would bias our measurements. The choice of our bins was to ensure that there is enough signal to provide a good spectral fit. We divided the brightness region into seven bins. Three of the seven bins were chosen to have the size of 163$^{\prime\prime}$ x 40$^{\prime\prime}$. However, the bins around the edge (four bins) were made to be smaller (163$^{\prime\prime}$ x 13$^{\prime\prime}$ - see Figure~\ref{figa2163_xray} - top-right panel). Higher resolution spatial binning is ideal for minimizing contamination due to bright emission just inside the edge and closely probes the shock \citep{Rus10}.
\begin{table*}
\begin{center}
\caption{Results from the fit of the temperature profile across the edge and in the perpendicular direction. The errors are within the 68\% confidence limit.}
\label{tabletemp}
\begin{tabular}{c  c  c  c  c  c|}  \\ \hline
\multicolumn{3}{c}{Shock Profile} &  
\multicolumn{3}{c}{Perpendicular Profile}  \\ \hline
Temperature (keV) & Distance (kpc) & Chi-squared  & Temperature (keV) & Distance (kpc) & Chi-squared  \\ \hline
 13.3$\pm$1.0 & -215.3 & 760.62: 768 PHA bins   & 12.3$\pm$1.4 & -452.3  &  378.395: 384 PHA bins \\ 
 13.0$\pm$0.8 & -81.2 &  1006.37: 982 PHA bins  & 13.1$\pm$1.2 & -301.5  &  605.636: 646 PHA bins \\    
 10.1$\pm$0.5 & 53.0  &   948.70: 1020 PHA bins & 12.4$\pm$0.7 & -150.8  &  820.300: 975 PHA bins \\ 
 11.8$\pm$1.4 & 141.2 &   373.84: 444 PHA bins  & 10.3$\pm$0.4 &  0.0    & 1168.240: 1175 PHA bins \\ 
 10.3$\pm$1.0 & 185.1 &   342.18: 402 PHA bins  & 11.3$\pm$0.5 &  150.8  & 1023.521: 1086 PHA bins \\ 
  4.1$\pm$0.3 & 229.4 &   548.62: 373 PHA bins  & 11.8$\pm$1.0 &  301.5  &  575.784: 684 PHA bins \\ 
  4.3$\pm$0.2 & 273.6 &   422.96: 316 PHA bins  & 12.5$\pm$1.6 &  452.3  &  269.328: 369 PHA bins  \\
\hline            
\end{tabular}
\end{center}               
\end{table*}
\subsection{Temperature Profile across the Edge} 
We thus obtained a temperature profile along the main axis of the cluster to determine the gas temperature jump across the edge. 
%to characterize edges as shocks
%
As a check, we also extracted spectra in the area shown in the bottom-right panel of Fig.~\ref{figa2163_xray}, using the same method described above, to obtain the temperature variation behaviour in the northwest - southeast perpendicular direction. The temperatures and the profiles are shown in Table~\ref{tabletemp} and in Fig~\ref{temp_prof} - left panel, respectively, and the errors are approximately 7\% -20\%. \\
We see a sudden drop in temperature (Fig.~\ref{temp_prof}) at the location of the brightness edge shown in Fig.~\ref{figa2163_xray}, where we argue there is a shock, which occurs at $\sim$ 200 kpc ($\sim 60^{\prime\prime}$) from the cluster centre. The temperature drops by more than 6 keV between before and after the edge. However, in the case of the perpendicular profile, we observe temperatures between 10.283 keV and 13.124 keV with no sudden drop in temperature on the galaxy cluster edges. This implies that the behaviour we observe in the area of the edge is specific and is found only on the brightness edge of this galaxy cluster.  Thus, the temperature profile confirms that the brightness edge is a shock front. The temperature jump of the right sign across the brightness edge leads to an approximate value of the ratio of the post-shock to pre-shock temperatures of 2.5$\pm$0.3. After applying the Rankine-Hugoniot jump conditions across the shock \citep{Lan59}, and assuming monoatomic gas with $\gamma = 5/3$, we determine the Mach number:
\begin{equation}
M=\bigg( \frac{(\gamma+1)^2(T_{2}/T_{1}-1)}{2\gamma(\gamma-1)} \bigg )^\frac{1}{2}
\end{equation}
to be $2.2\pm0.3$.
This value of the Mach number is a lower limit since for the plane shock the maximum of temperature should be just behind the shock (see Fig~\ref{temp_prof} - left panel).  \\
The drop in temperature near the centre of the cluster in both profiles (Fig~\ref{temp_prof} - left panel) is due to the presence of the cool core reported by \cite{Bou11}. The cool core is located at approximately $\sim$50 - 100 kpc from the centre, in the SW direction.  \\
In Fig.~\ref{temp_prof}-middle panel, we show a simple best-fit shock model on our temperature profile. To model the temperature profile we used a simple step function 
\begin{equation*}
T(r)=\left\{ 
 \begin{array}{rl}
  T_0 & r \le r_{j}, \\
  T_1 &  r > r_{j}.
 \end{array} \right\}
\end{equation*}
where $r_j$ is the location of the jump discontinuity. From the fit we obtained the step amplitude (difference between temperature values before and after the discontinuity caused by the shock) which is 7.0$\pm$ 0.7 keV, leading to a jump in temperature of 2.7$\pm$ 0.6. The reduced Chi-squared value is $\chi^2 / d.o.f = 3.2$. This jump leads to the Mach number $M= 2.3\pm$ 0.5, in good agreement with the above value of the mach number.  
\begin{table*}
\begin{center}
\caption{The properties of ACO2163 and the details of KAT-7 observations.}
\label{tab:prop}
\begin{tabular}{c c c c c}
\\
\hline\hline
Cluster Name & z &    RA (J2000) & DEC (J2000) &  Diameter (arcmin) \\
	    &     &    (h m s)   &  (min s arcs) & (at 5R500: X-ray)  \\   \hline
ACO2163    & 0.20    &  16 15 34.1 &  -06 07 26    &   37.81	     \\
\hline
\end{tabular}
\end{center}
%\end{table*}
%
%
%
%\begin{table*}
%\centering
%\caption{Details of KAT-7 observations of ACO2163.}
%\label{tab:log}
\begin{tabular}{c c c c c c c c}
\\
\hline\hline
Cluster Name   &  Freq.       &   BW   & Observation Date & Observing time  & Antennas &   FWHM, p.a.          & rms             \\  
	       &    (MHz)    &  (MHz) &                  &        (hours)  &            &  ($^{\prime\prime}$ x$^{\prime\prime}$), degree  & (mJy beam$^{-1}$)   \\ \hline
ACO2163	   &  1826.6875	   & 234765 &  05-Oct-2012 & 4.48  & 7                          & 226$\times$162, 153           &  1.1              \\
\hline
\end{tabular}
%\end{center}
\end{table*}
\subsection{Luminosity Profile across the Edge}
We looked at the luminosity profile in the region around the presumed shock to check for a sudden decrease in luminosity. To this end, we used bin sizes (165$^{\prime\prime}$ x 29$^{\prime\prime}$) which provided enough signal to extract spectra and obtain the luminosity. 
We chose the smallest possible window providing converging X-ray parameter estimates to have as many measurements as possible along the direction of the presumed shock.  \\
To obtain the luminosity, we used the {\it mekal} model using temperature, H density and abundance as free parameters. Since we know the redshift of Abell 2163, we set its value to 0.203 and we let the other 3 parameters free. The luminosity profile is shown in Fig.~\ref{temp_prof}-right panel (solid line). \\
The brightness edge is clearly visible in the luminosity profile as a sudden drop in luminosity by a factor of $\sim$5.2. The luminosity drops by $\approx 2.8 \times 10^4 ergs/s$. The jump in luminosity occurs at a distance of $\sim 200$ kpc from the cluster centre, which is the same location where we observe a jump in temperature. We also looked at the luminosity profile in the sector that is perpendicular to the SW direction where the shock is located (dashed line). We observe no luminosity jump in this direction. We thus conclude that the luminosity jump is associated to the shock. \\
Since the X-ray luminosity $L_x \propto n^2 T^{-3/4} V$ for the XMM band we are using \citep{Bas10}, hence we can re-write this relation as
\begin{equation}
{\frac{L_x^{post}}{L_x^{pre}}} = \Big( {\frac{n^{post}}{n^{pre}}} \Big)^2 \Big({\frac{T^{post}}{T^{pre}}}\Big)^{-3/4} \times 2.3 
\end{equation}
after re-scaling the volume of smaller boxes that were used for temperature to that of luminosity boxes. This leads to
\begin{equation}
\frac{n^{post}}{n^{pre}} = \Big( \frac{L_x^{post}}{L_x^{pre}} \Big)^{1/2} / \Big [ \Big( {\frac{T^{post}}{T^{pre}}} \Big)^{-3/8} \times (2.3)^{1/2} \Big ]
\end{equation}
where $n$ is the mean gas density and $V$ is the gas volume, 'pre' and 'post' refer to pre-shock and post-shock regions, respectively. 
Using the previous values for the temperature jump (2.5$\pm$0.5) and the estimated luminosity jump (5.2 $\pm$ 0.4) we obtain an approximate value for the jump in density, $\frac{\rho_1}{\rho_2}$, which is 2.2 $\pm$ 0.5. The Mach number of the shock front was then determined by employing the Rankine-Hugoniot shock relation 
\begin{equation}
M=\left (\frac{2\frac{\rho_1}{\rho_2}}{(\gamma +1)-(\gamma -1)\frac{\rho_1}{\rho_2}} \right )^{1/2}
\end{equation}
and was found to be 1.9 $\pm$ 0.4, which is consistent with the Mach number obtained for the temperature.  \\
The jumps are calculated without performing a deprojection of the temperature profile. The possible effect is that the Mach number might be underestimated.
\section{Radio Observations with The Karoo Array Telescope (KAT-7)}
\label{sec:robs}
\begin{figure*}
\centering
\includegraphics[scale= 0.5,width=130mm,height=140mm,bb=0 150 480 730, clip=true]{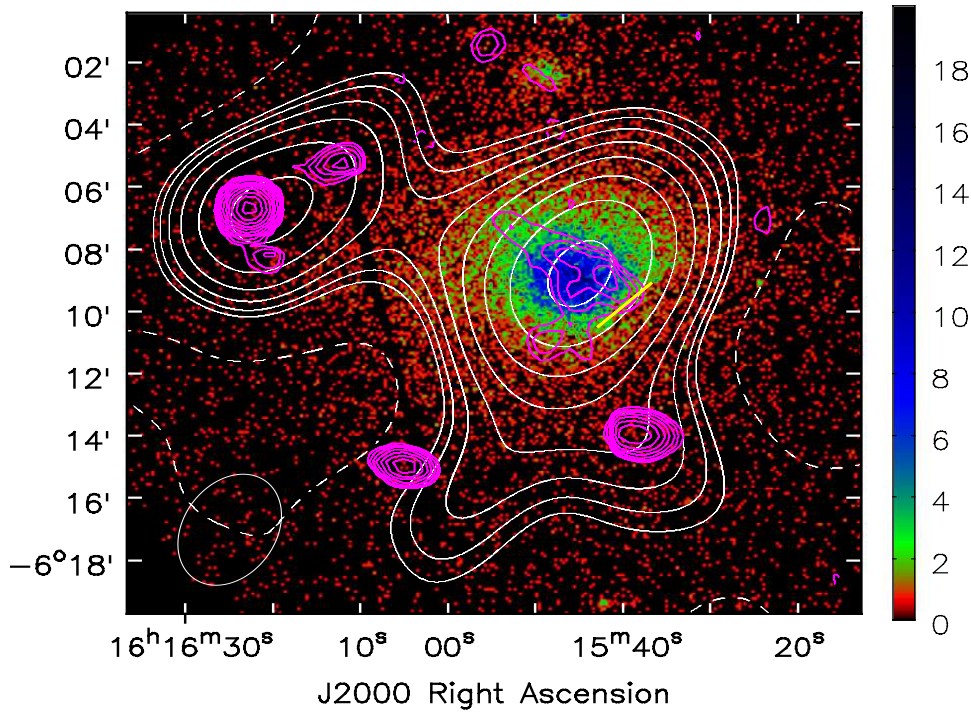}
\vspace{-4.5cm}
\caption{KAT-7 radio contours of ACO2163 (white) after subtraction of background sources in the field of the cluster are overlayed on the XMM-Newton image. The radio contours of the halo in NVSS are overlayed on the image in magenta. The restoring beam of the radio image is 226$^{\prime\prime}$ $\times$ 162$^{\prime\prime}$ (PA=153 deg), and the noise level in the image plane is 1.1 mJy/beam. The contours start at 3.3 mJy/beam and then scale by a factor of $\sqrt2$. The restoring beam is in the bottom-left hand corner. NVSS contours start at 1.6 mJy/beam and also scale by a factor of $\sqrt2$. The yellow line indicates the location of the shock.}
\label{fig:aco2163_nvss}
\end{figure*}
ACO2163 was observed in 2012 with the decommissioned seven-dish KAT-7 array, an engineering testbed for the MeerKAT which is South Africa's pathfinder for the Square Kilometer Array (SKA).  \\ 
The KAT-7 telescope has been used to detect HI in nearby galaxies (Carignan et al. 2013), to study extended radio haloes in galaxy clusters (Riseley et al. 2015; Scaife et al. 2015) and to study time-variability of radio sources (Armstrong et al. 2013), and we refer to these papers for the technical specifications of the telescope. \\
ACO2163 is one of the eight galaxy clusters in our sample that were observed at 1.83 GHz using the KAT-7 array. The sources were selected based on their extent, i.e. they had a diameter that is more than 20$^{\prime}$, from a list of clusters that are observable in the Southern Hermisphere.  The data were obtained with a total observing bandwidth of 235 MHz divided in 601 channels. The properties of ACO2163 and the observation information are summarized in the compound Table~\ref{tab:prop}. \\
The data were reduced and calibrated using the NRAO Common Astronomy Software Applications (CASA) (McMullin et al. 2006), which is the standard data reduction package that is used for the reduction of the KAT-7 data. \\
Before the calibration, bad data and RFI were removed within CASA, and the 601 channels were averaged to 9, each having a width of 25 MHz in order to reduce the size of the data set. The interval width for time averaging was 25s, and 64 channels were averaged to output each of the 9 channels. 
The choice for these values was to ensure that time/frequency smearing are minimized. Flux calibrator source PKS 1934-638, and the phase calibrator PKS 1621-115, were used to calibrate the ACO2163 data in flux and in phase, respectively. The flux density calibrator source was also used to do bandpass calibration. \\
After a number of self-cal procedures which were applied in order to reduce residual phase variations, deconvolution was achieved using the CLEAN algorithm in CASA. The images were produced using a Briggs robustness parameter of 0.5.  This value (Briggs=0.5) has the advantage of minimizing noise while giving a better resolution with respect to natural weighting, allowing for good point source detection in the resultant images. Point sources in the background were identified and subtracted in the uv data using the task UVSUB before the search for diffuse emission was done. Only sources above the confusion limit of 1 mJy/beam were subtracted.
{\it nterms=2}, which is the number of terms in the Taylor polynomial, was used to model the frequency dependence of the sky emission.  \\
A multi-scale CLEANing algorithm in CLEAN was used in order to detect diffuse, extended structures on both smaller and large spatial scales. 
In order to pick up the diffuse emission, the final background source-subtracted image was produced by means of natural weighting.
\begin{figure}
\centering
\includegraphics[scale=0.6]{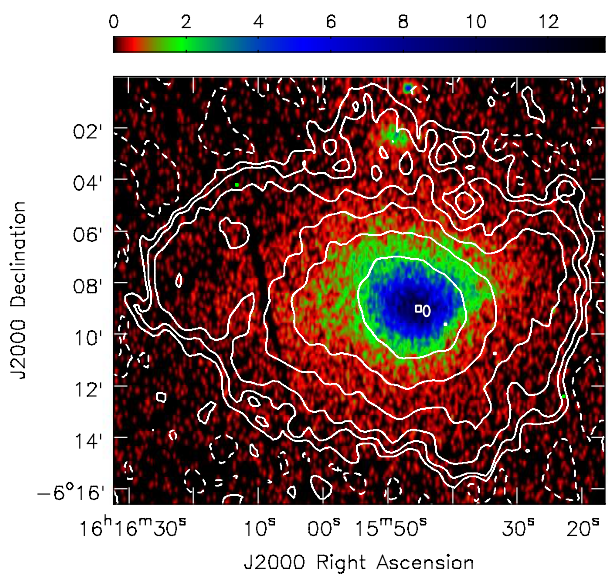}
\caption{Radio contours of the halo in VLA at 1.4 GHz, with resolution of 45$^{\prime\prime}$ $\times$ 60$^{\prime\prime}$, are overlayed on the XMM-Newton X-ray image. The $\sigma$ noise level in the radio image plane is 0.03 mJy/beam. Contours are at -3, 3, 6, 12, 24, 48, 96 mJy/beam times the noise level. The units in the colour bar are mJy/beam. The radio image is from \protect\cite{Fer01}.}
\label{fig:aco2163_vla_low}
\end{figure}
\section{Radio Analysis and Results}
\label{sec:radio}
We have detected the ACO2163 radio halo and the extended north-eastern relic source in the KAT-7 data (Figure~\ref{fig:aco2163_nvss}). The two sources seem to be connected by a wide bridge. The radio halo has a total size of $\sim$9.5$^\prime$ and a total flux density $S_{1.83GHz} = 93 \pm 1.3$ mJy. In the same figure we also show NVSS contours overlayed on the XMM-Newton X-ray image, as well as the approximate location of the shock which is indicated by a straight yellow line. The elongation in the north-south direction is possibly due to the emission associated with the two point sources seen in the NVSS contours. The flux density in the KAT-7 image is lower but close to the expected value of $\sim$103 mJy obtained by scaling the 1.4 GHz flux to that at 1.83 GHz using the spectral index across the 20 cm band which is about 1.6 (Feretti et al., 2001). The discrepancy could be due to some negative residuals, as suggested by the negative bowl in the bottom left in Figure~\ref{fig:aco2163_nvss}, where there is extended emission detected from the VLA.   \\
Since the KAT-7 observations are of poor resolution, our discussion of the flux results for the halo are limited, as well as those of the extended north-eastern relic source which will follow later.  \\
In addition to the KAT-7 data, we obtained ready-to-use radio halo and spectral index maps of \cite{Fer01,Fer04} that we utilised to read out spectral and flux profiles. The radio maps were obtained at 1.4 GHz and 0.3 GHz with the VLA by \cite{Fer01,Fer04}. The high and low resolution images of the radio halo at 1.4 GHz are shown in Figures~\ref{figa2163_xray} and~\ref{fig:aco2163_vla_low} as contours overlaid on the X-ray image, and show the detection of a giant radio halo, with a total extent of $\sim$11.5$^{\prime}$ (2.3 Mpc) and an elongation in the E-W direction. After the subtraction of point sources, the total fluxes in the halo were found to be $S_{1.4GHz} = 155 \pm 2$ mJy and $S_{0.3GHz} = 861 \pm 10$ mJy, which led to an average spectral index $\alpha^{1.4}_{0.3} = 1.18 \pm 0.04$.
 \\
In the low resolution images of the VLA and KAT-7, the shock front is further away from the edge towards the cluster centre, well within the radio halo region of a dynamically disturbed cluster. 
Given its short baseline spacings, the KAT-7 sees all the scales and is picking up more diffuse emission from the cluster which could explain the large size of the halo.
In the high resolution VLA image in Figure~\ref{figa2163_xray}, the difference between the location of the shock and that of the halo edge is roughly about 2 beams. This location of the shock front could suggest a connection between this shock front and the diffuse emission at the halo edge.

\subsection{Spectral Index Analysis}
\begin{figure}
\centering
\includegraphics[scale=0.9]{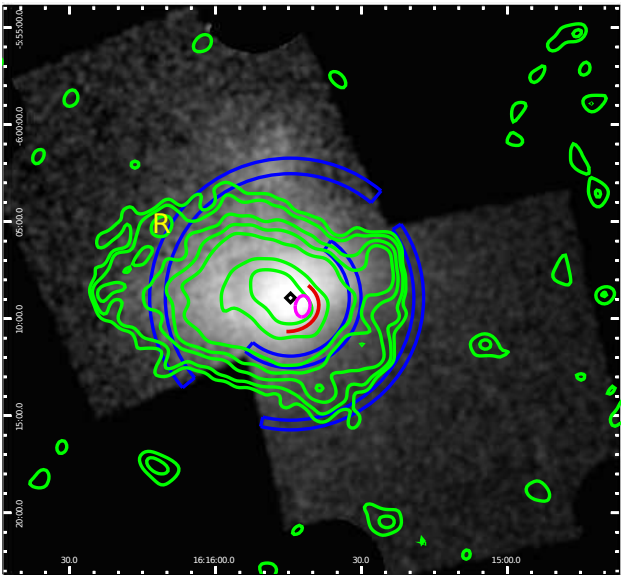}
\caption{This figure, obtained from \citet{Thol18}, shows the positions of the shock fronts they detected in A2163 (blue), and the position of our newly detected shock front (red) near the approximate position of the cool core (magenta ellipse). The black diamond represents the X-ray emission peak.}
\label{fig:shock_locations}
\end{figure}

The spectral index map was obtained by comparing the 90 cm and 20 cm images produced with the same beam and cellsize.
We think that the flux is not missed at 20 cm and therefore the spectrum is not affected by the slightly different u-v spacings for the following reasons: (i) diffuse emission at the 2 frequencies has similar extent,
(ii) the shortest baselines at 90 cm are provided by very few interferometers,
(iii) the spectral index map has values consistent with the total spectral index, and
(iv) the 20 cm map is much more sensitive than the 90 cm one (see details in \cite{Fer04}) \\
As expected from electron re-acceleration models, regions influenced by the shock (and turbulence) should show spectral flattening, indicating the presence of more energetic radiating particles and/or a larger value of a local magnetic field strength \citep{Fer01}. \cite{Fer04} identified two regions where they saw evidence of flat spectra in the spectral index map: the vertical region crossing the cluster centre (in the N-S direction) to a distance of $\sim$300$^{\prime\prime}$ from the cluster centre, and the western halo region which was observed to be much flatter than the eastern region.
This is the same side of the cluster where the new shock front is located, as well as the two other shocks which have been detected recently \citep{Thol18}, one at $\sim700$ kpc and the other at $\sim1300$ kpc from the cluster centre (see Figure~\ref{fig:shock_locations}). 
The flat spectra in the N-S region were interpreted as an indication of strong dynamical activity due to a merger occuring in the E-W direction. 
We investigate if some of the flattening observed in the south-western region could be due to the activity of the new shock. 
\begin{figure*}
\centering
\includegraphics[scale=0.5,width=130mm,height=110mm]{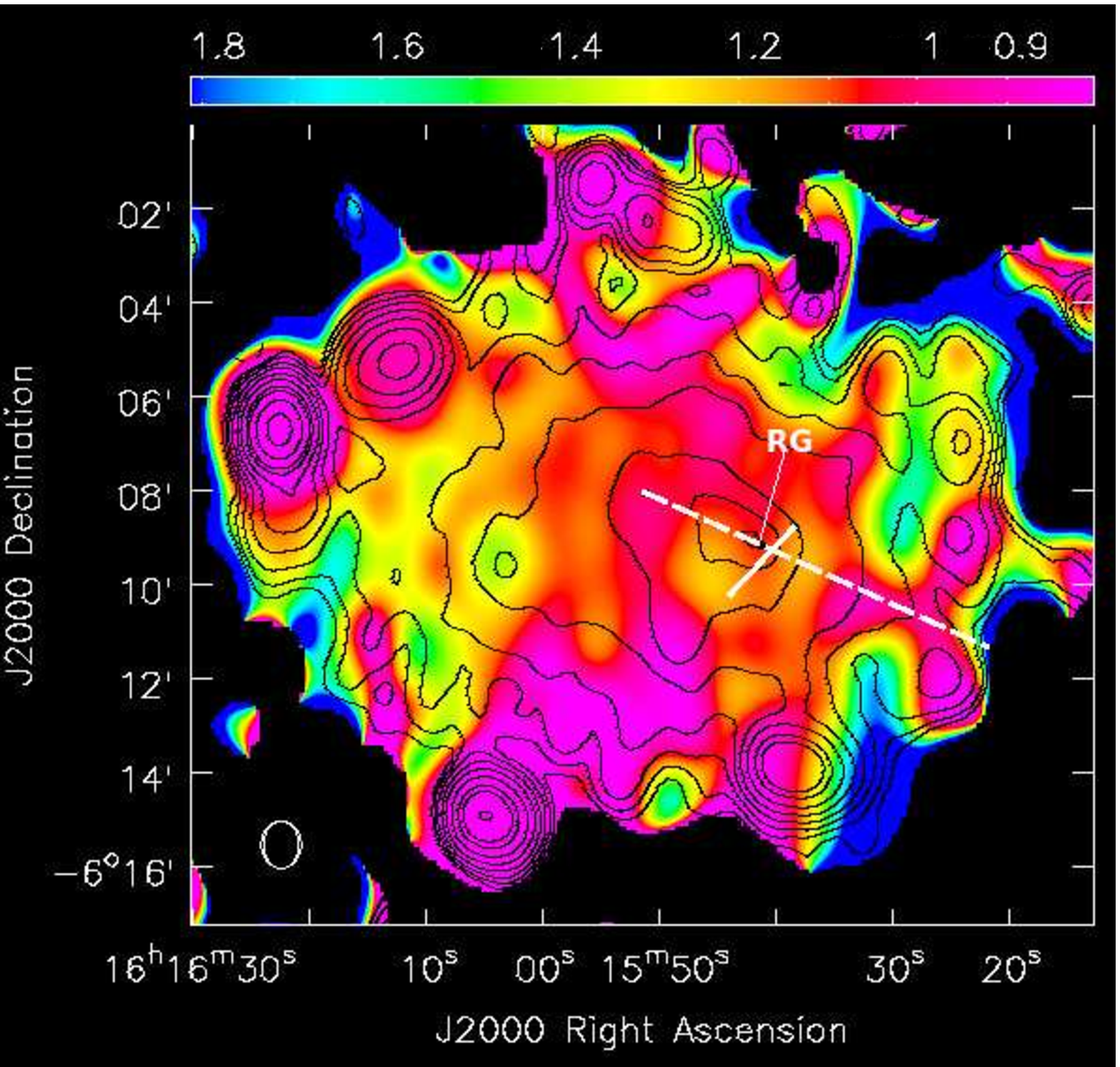}
\includegraphics[scale=0.5,width=75mm,height=48mm]{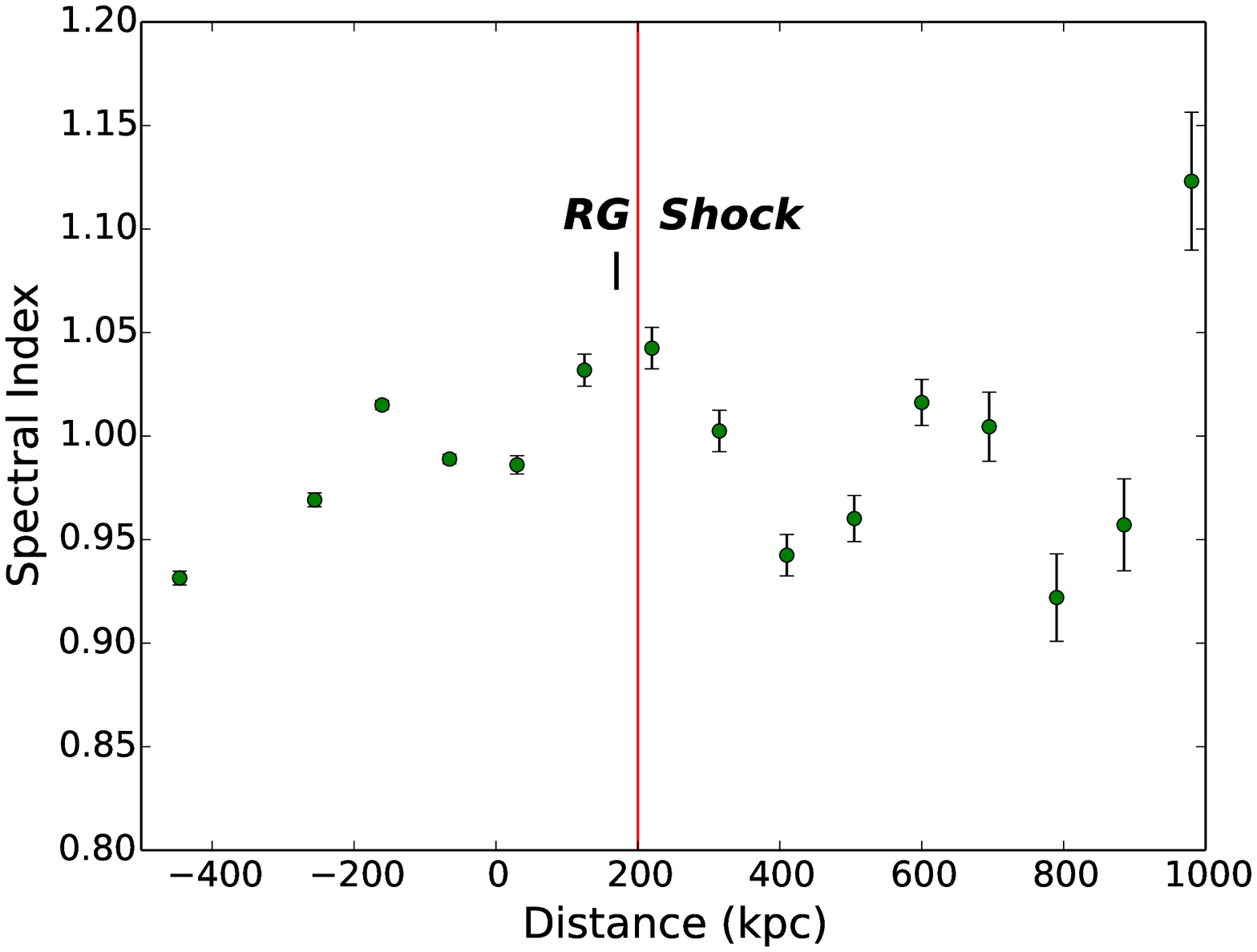}
\includegraphics[scale=0.5,width=75mm,height=48mm]{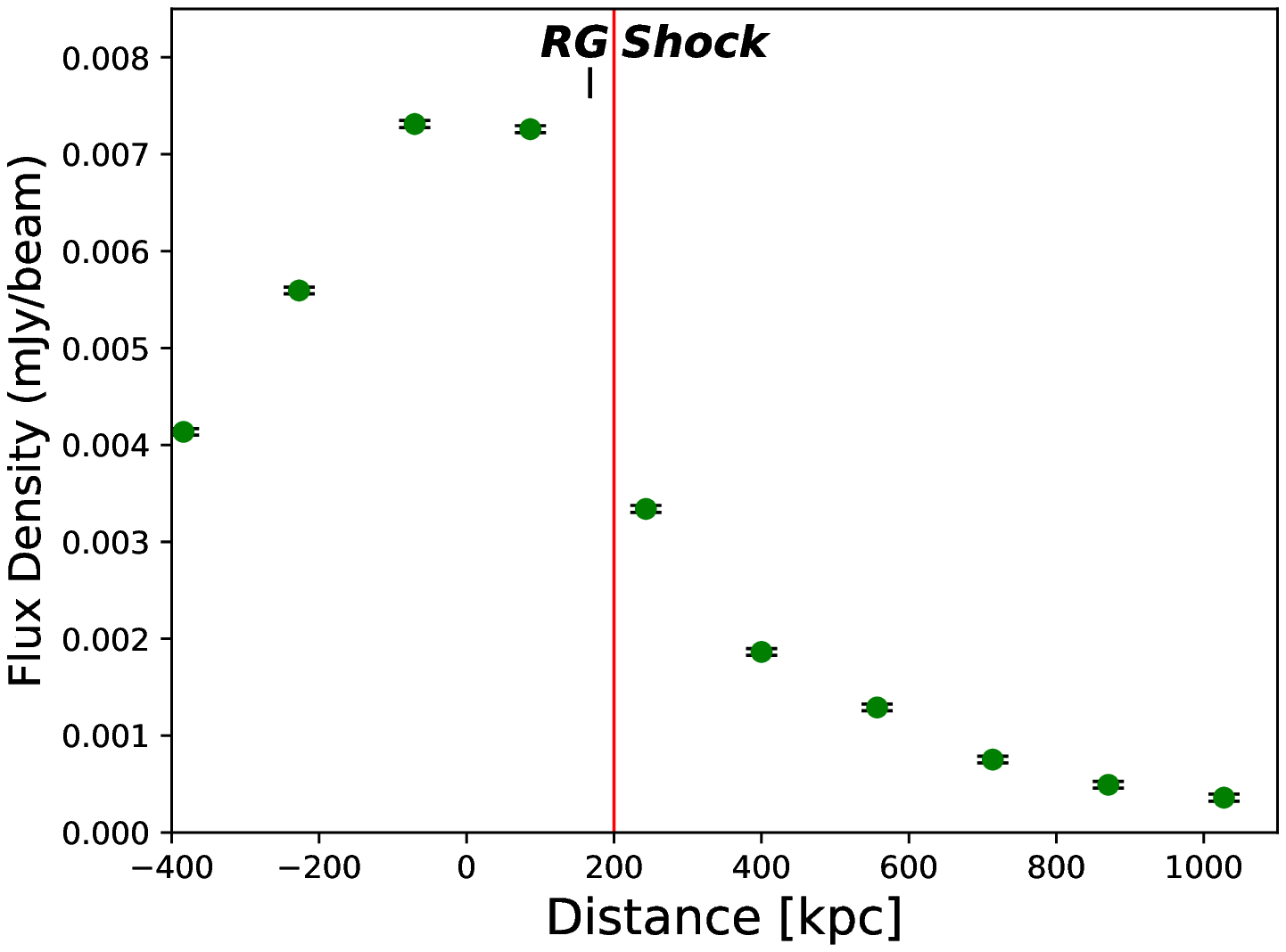}
\caption{Top panel: Radio contours of the halo in VLA at 1.4 GHz, with resolution of 45$^{\prime\prime}$ $\times$ 60$^{\prime\prime}$, are overlayed on the colour-scale spectral index image between 0.3 GHz and 1.4 GHz. The image is obtained with a resolution of 60$^{\prime\prime}$ $\times$ 51$^{\prime\prime}$. The contours are at -3, 3, 6, 12, 24, 48, 96 mJy/beam times the noise level. The location of the shock is represented by a solid white line and that of the radio galaxy is also indicated. The spectral index image is from \protect\cite{Fer04}. Bottom panels: 1-dimensional slice, Left: of the spectral index image showing the spectral index distribution along the direction indicated by the dashed line segment that passes through the indicated location of the radio galaxy in the spectral index image. The origin of the distance scale starts at the approximate cluster centre at RA=16h15m51.2s, -06d08m01.6s. The locations of the radio galaxy and the shock are indicated. Right: showing the flux density distribution in the low resolution VLA radio image along the same dashed line segment shown in the spectral index image. The vertical line indicates the location of the radio edge.}
\label{fig:spinx-vla-prof}
\end{figure*}
Furthermore, we note the presence of a radio galaxy (J1615-061) behind the shock (Figure~\ref{fig:spinx-vla-prof}) at RA=16h15m41.3s, DEC= -06d09m08s, which was first reported by \cite{Fer01} (the source is labelled T2 in their Figure 2). If the radio galaxy is in the volume of the halo, the fossil radio plasma would affect the spectral index in the region of the halo, leading to a steep spectrum \citep{Kan15, Kan16}, or a flat spectrum if shock re-acceleration does not stop after the shock crosses the fossil plasma cloud, or there is ongoing merging activity in the region. \\
To investigate if the flatter spectra in the S-W region could be due to the activity of the shock, and if there is any possible connection between the radio galaxy fossil electrons and the shock, we produced spectral index and flux radial profiles along the sector represented by the white dashed line in the spectral index image (Figure~\ref{fig:spinx-vla-prof}), emerging from the turbulent region near the cluster centre, passing through the radio galaxy location, to the radio halo edge. \\
Both the spectral index and flux profiles were obtained by averaging spectral index values from the spectral index image and flux values using the low resolution halo image, respectively, within small sectors along the direction indicated by the dashed line. In both cases the sector area was made to be roughly equal to that of the beam. 
The spectral index profile (Figure~\ref{fig:spinx-vla-prof} - bottom-left panel) does not show any significant spectral variation but shows a globally flat spectral index. \\
We varied the extraction region slightly up and down by a few beams to test whether the spectral trends hold. The spectral trends observed in Figure~\ref{fig:spinx-vla-prof} - bottom-left panel - do not change after this exercise.  
Though we do not observe significant flattening at the specific location of the shock as expected by re-acceleration models, the spectral index is globally flat on this side of the halo, indicating the presence of more energetic radiating particles.   \\
Another example of a Mach number $\sim$2 shock which has not shown any evidence of spectral flattening at the shock location is observed in the cluster ACO665 \citep{Mar01, Fer04}. Instead, progressive steepening with distance on the south-eastern side of the cluster, where the shock is located, is observed and suggests that this cluster region is undisturbed by the merging processes, and that shocks in major mergers are probably weak for particle acceleration \citep{Gab03}.
It is important, however, to stress that though we do not detect any significant flattening at the location of the shock, the progressive steepening seen in ACO665 is not observed either on the western side of the cluster ACO2163 where the shock is located. Instead, we observe a globally flat spectral index. This could be a result of the recent two-component interaction, and the westward motion of a stripped cool core/`bullet' \citep{Bou11}, on this side of the halo which we argue has produced the $M=2.2$ shock. The relatively globally flat spectral index could indicate the presence of more energetic radiating particles due to shock activity in this region as already indicated \citep{Thol18}.  \\
The bottom-right panel in Figure~\ref{fig:spinx-vla-prof} shows the flux density distribution along the dotted line segment passing through the radio galaxy region in the spectral index image. The flux density peaks near the location of the radio galaxy core just behind the shock front, and rapidly declines on the side of its lobe \citep[see][]{Fer01}, up to $\sim$250$^{\prime\prime}$ from the core location. The sudden decline in radio flux corresponds with the radio edge.
%The peak is the combined flux of the halo and the radio galaxy.
%
\subsection{The North Eastern Radio Relic}

The shock front reported in this paper is embedded in the halo and as a result we do not observe any relic source coincident with the shock.
A diffuse source on the N-E side of the halo in ACO2163 claimed to be a relic, however, was detected by Feretti et al. 2001 (labelled D3 in their Figure 2). Another emission, D4, was seen south of D3, and no connection between the two sources was established. Feretti et al. 2001 ruled out the possibility that these two sources are linked with the strong eastern radio source, J1616-061 (hereafter source S), located at RA=16h16m22s and Dec=-06d06m34s (see also overlaid NVSS contours in Figure~\ref{fig:aco2163_nvss}). Later observations by Feretti et al. 2004 confirmed D3 (which is labelled 'R' in their Figure 2) as a relic source, with a spectral index $\alpha^{1.4}_{0.3} = 1.02 \pm 0.04$. \\
After attempting to subtract source S, located $\sim$9$^{\prime}$ (1.8 Mpc) from the cluster center in our KAT-7 image, we detect large-scale diffuse emission which encompasses both D3 and D4. The size of this source is $\sim$5$^\prime$ ($\sim 993$ kpc).    \\
The measured flux of the relic before subtraction of the point source S is 55 mJy. The subtracted clean component of S contributes a flux density of only $S_{1.83GHz} = 33$ mJy in KAT-7. So the resulting flux of the relic source should be $\sim$22 mJy after subtraction. However, after subtraction we measure a flux density of 42.6 mJy for the relic, $\sim$50\% more than what we expect. 
The reason for this discrepancy could be that the point source was picked up using rhobust = 0.5 which gives high resolution. After subtraction in the uv data the final point source subtracted image was produced with weighting = natural. So the point source in the weighting = natural image would be more extended than in the rhobust = 0.5 image, thus leaving some residual emission after subtraction. \\
Feretti et al. 2004 measured a value of $S_{1.4GHz} = 18.7$ mJy for the relic source, and our expected value of $S_{1.83GHz} = 22$ mJy suggests that the relic source in ACO2163 is more extended due to the KAT-7 high sensitivity, with possibly more flux density at 1.4 GHz than was previously observed (i.e. less flux was possibly picked up at 1.4 GHz due to the high resolution of the long-baseline interferometer). \\
Feretti et al. 2004 suggested the possibility that source S is a tailed radio galaxy, extended toward the south. If this is the case then that would mean our relic emission is also contaminated by the tail emission of this source (source D), which would also explain the high flux value that we measured for the relic. Though it is not possible to determine the accurate flux value for this source in KAT-7, it is reassuring to notice that the location of the $M=1.7$ shock front of \cite{Thol18} coincides with this relic source. \\
In order to establish the exact extent of the relic source in ACO2163, new observations are required and the spectral index analysis of source S is necessary to see if there is any progressive spectral steepining from the centre of the source to the end of the potential tail toward the south. 
\section{Discussion}
\label{sec:disconc}

The main goal of this paper is to present evidence for the existence of a shock front in ACO2163, located at $\sim$60$^{\prime\prime}$ ($\sim$200 kpc) from the cluster center in the X-ray map,
%located at $\sim$175$^{\prime\prime}$ ($\sim$600 kpc) from the cluster center
 on the south-western region of the cluster. We have derived the Mach number associated with this shock from the direct measurement of a gas temperature jump. The temperature jumps from 4.1 keV to 10.3 keV by a factor of 2.5$\pm$0.3 at the location of the shock front. This temperature jump corresponds to a Mach number $M=v_{s}/c_{s}$ $= $2.2$\pm$0.3, where $c_{s}$ is a sound speed of $\sim$1021 km s$^{-1}$. The shock velocity $v_{s}=Mc_{s}$ has been determined to be $\sim$2238 km s$^{-1}$. \\
A detection of a shock front in ACO2163 supports the results of \cite{Bou11} which have revealed I) the `bullet' nature of the supersonic infalling substructure in ACO2163, II) the westward motion of a stripped cool-core/`bullet' embedded in the hotter atmosphere of ACO2163-A,  and III) adiabatic compression of the gas behind the `bullet' due to shock heating. These observations point to a shock front being present in the inner regions of ACO2163, which we argue we have detected. The results of \cite{Bou11} also showed the similar nature of ACO2163 to that of the `Bullet' cluster where a shock front that was preceding a 'bullet' was detected \citep{Mar02}. \\
Given the location of the `bullet' which is at a projected distance of 290 kpc from the main cluster \citep[see][]{Bou11}, and assuming that the subcluster moves at the shock velocity, this velocity implies that the `bullet' crossed the main cluster nearly 0.1 Gyr ago.  \\
Our value of the Mach number is a typical value for a shock front in merging clusters (\citep[see e.g.][for ACO520]{Mar05}, \citep[for ACO754]{Mac11}, or \citep[for CIZA2242, ACO3667, and ACO3376]{Aka13}), and since shock fronts are routinely observed in merging clusters, the detection of a shock in ACO2163 supports the results \citep[e.g.][]{Bou11} which show the merging nature of this cluster.  \\
Radio observations of a number of merging galaxy clusters provide evidence for the presence of relativistic electrons and magnetic fields in the intracluster medium through the detection of synchrotron radio emission in the form of Mpc-scale sources called radio halos \citep[e.g.][]{Fer12, Bru14}.
The relativistic electrons can be (re)accelerated by turbulence and shocks in the ICM, and the process of shock re-acceleration can be efficient for shocks with M$>$2 \citep[e.g.][]{Mac11}, even though its efficiency is uncertain, but is thought to be small in the innermost cluster regions, and increases towards the virial radius \citep{Vaz11}.  \\
If one were to assume direct acceleration via {\it Fermi}  mechanism of relativistic electrons by the M $\sim$ 2.2 shock that we detect in ACO2163 \citep[e.g.][]{Blan87, Mar05}, the spectral index in the shock region (accounting for down-stream losses due to inverse Compton and synchrotron energy losses) should be $\alpha = (p-1)/2 + 0.5$ $\sim$ 1.53 ($\alpha = \alpha_{inj} + 0.5$ and $\alpha_{inj} = (p-1)/2$ $\sim$ 1), where $p$ is the slope of the energy spectrum of the electrons generated by the shock, and is related to the Mach number of the shock through
\begin{equation}
p=2\frac{M^{2} + 1}{M^{2}-1}
\label{eqn:1}
\end{equation} \citep[e.g.][]{Blan87}.
This spectral index value, which suggests a steep spectrum, is not consistent with the total integrated spectral index $\alpha^{1.4}_{0.3}=1.18\pm$0.04 obtained by \cite{Fer04} and with that of $\sim$1 observed around the region of the shock (Fig~\ref{fig:spinx-vla-prof}) which both predict a value of the Mach number of $\sim$3.5 or higher. \\
The discrepancy between the observed spectral index and that predicted by the shock (or between the corresponding Mach numbers) has been observed in another merging cluster 1RXS J0603.3+4214 (or the Toothbrush cluster - see \cite{vWe12, Ogr13}, see also \cite{Col17}).  \\
The mixing of the emission, which can be an important factor that determines the spectral shape \citep{vWe12} has been offered as one of the possible explanations for the observed discrepancy. The emission at the location of the radio galaxy where we detect a peak flux (Fig~\ref{fig:spinx-vla-prof}) is possibly a mixture of the radio galaxy emission and that from the radio halo. This mixing of emisssion suppresses spectral curvature and pushes back the radio spectra to power-law shapes. \\
It has also been suggested that projection effects could smooth out temperature discontinuities across the shock front, leading to a low value of the Mach number \citep{Ogr13}. \\
In fact, \cite{Fer01} found a total spectral index of $\alpha(int)^{1.465}_{1.365} \sim 1.6$ for the ACO2163 halo, between two nearby frequencies, and pointed out that even allowing for large errors, this could be an indication of a spectral steepening above $\sim$1.4 GHz.  \\
We have noted the presence of a radio galaxy along the path of the shock which could have provided the fossil electrons, and this radio galaxy has a NVSS flux of 16.57 mJy, but there is no spectral information about this source that is available. However, from our analysis we note two things about this source: (i) The radio galaxy core region fits the global trend of a flat spectrum, with a spectral index value of $\sim$1.0 and (ii) the flux density peaks at the location of the radio galaxy core.  
If we assume that the electrons from the radio galaxy are injected with $\alpha_{inj}$ or $\alpha_{s}+3 \sim$ 1 (which is the spectral index at the location of the shock - see Fig~\ref{fig:spinx-vla-prof} - bottom-left panel), this injection spectral index in combination with DSA, would implicate a Mach number 
\begin{equation}
M_{radio}=\bigg (\frac{2 \alpha_{s}+3}{2 \alpha_{s}-1} \bigg)^{\frac{1}{2}}
\end{equation}
$\sim2.2$, 
where $M_{radio}$ is the Mach number from the radio observations for a steady planar shock. This value of the Mach number is in excellent agreement with that derived from the X-ray observations, in support of re-acceleration by DSA.  This could mean that the electron population injected at the shock with M $\sim$ 2.2 results in a steeper spectrum with $\alpha \sim 1.6$ at higher frequencies. \\
The evidence for a relatively globally flat spectral index in the S-W region of the halo where the shock is located, contrary to the expected radial steepening in undisturbed regions of a cluster, indicates the presence of more energetic radiating particles, due to both the shock and merger processes being active in this region, leading to the possible mixing between the shock accelerated and turbulence accelerated plasmas. And for this reason it is not possible to determine the contribution of the shock-accelerated plasma based on this study, and so the above result should be viewed with caution. Higher resolution data at a few different frequencies would be needed.  \\
Furthermore, although the shock may seem to be in the tail of the radio galaxy in 2D projected image, the radio galaxy and the X-ray shock may not overlap in 3D space. Since the radio galaxy looks undisturbed in \cite{Fer01}, the possibility that the X-ray shock is not interacting with the radio galaxy in ACO2163 should also be considered, i.e. there are projection effects. But since the radio halo should be approximately spherical and is very extended, the shock should occur within the radio emitting region.   \\
Our new detection has increased the number of shock fronts detected in this cluster to four, explaining why ACO2163 is one of the hottest clusters of galaxies.
The three shock fronts of \citep{Thol18} suggest that prior to the current 2-component interaction, a major merger event occured which resulted in turbulence acceleration which produced the radio halo, and the associated shock which produced the N-E relic.  

\section{Summary and Conclusion}
\label{sec:disconc-1}
In this paper we have presented a combined X-ray and radio study of the nearby galaxy cluster ACO2163. \\
The following are the main findings from our study and analysis:  \\
1. The XMM-Newton observations have confirmed the presence of a shock front south-west of the cluster centre in ACO2163. We have estimated the Mach number from the X-ray temperature ratio before and after the shock, and have found a value of M=2.2$\pm0.3$. This is a typical value for a shock front in a galaxy cluster, and supports the results showing the merging nature of ACO2163.  \\
2. We observe a globally flat spectral index on the south-western side of the cluster where the shock is located, indicating the presence of energetic electrons in this region.   \\
3. We find a discrepancy between the observed average spectral index and that predicted by the shock (or between the corresponding Mach numbers). The average spectral index in the region of the shock predicts a Mach number $M=3.5$ or higher, which is higher than that observed in most merging clusters.  \\
4. The mixing of the emission, most likely due to projection effects, can be an important factor that determines the spectral shape, and has been provided as a simple explanation for the discrepancy in the spectral index. Another explanation entails the smoothing out of the temperature discontinuities across the shock front by projection effects, which could lead to a low value of the Mach number.\\
5. In combination with DSA, a steep spectral index ($\alpha \sim 1.6$) which was observed for two nearby frequencies in ACO2163, implecates a Mach number $M=2.2$, in good agreement with our X-ray results.  \\
The case of the shock in ACO2163 is a unique example since, in combination with re-acceleration of pre-existing electrons which is obvious, turbulent acceleration seems also important behind the shock (within the halo).  \\
The KAT-7 results seem to suggest a more extended radio emission at the relic source location. This is due to the large beam and enhanced sensitivity of this instrument. With the MeerKAT we will be able to detect halo and relic sources with radio power approximately ten times better when compared to existing arrays, and this will enable a more detailed study of the interaction between the shock front and the radio plasma.
\section{Acknowledgements}
NM would like to thank Thomas Reiprich, Sofia Tholken, Kaustuv Basu and Reionout van Weeren for the insightful discussions which helped improve the quality of this paper. This work is based on the research supported by the National Research Foundation of South Africa (grant number 111735). The KAT-7 telescope is operated by the South African Radio Astronomy Observatory, which is a facility of the National Research Foundation, an agency of the Department of Science and Innovation.
\section{Data availability}
Data is available on request from the authors.
\bibliographystyle{mnras}
\bibliography{mnras_shock}
%
\iffalse

\fi

\end{document}